\documentstyle[12pt]{article}
\def\ii{\'{\char'20}}

\begin{document}

\def\r{\rightarrow}
\def\G{\Gamma}
\def\f{\phi}
\def\p{\partial}
\def\ii{\'{\char'20}}
\def\brr{\begin{array}}
\def\err{\end{array}}
\def\bea{\begin{eqnarray}}
\def\eea{\end{eqnarray}}
\def\beq{\begin{equation}}
\def\eeq{\end{equation}}
\def\bs{\bigskip}
\def\tr{\mbox{Tr}\, }
\def\ni{\noindent}
\def\wt{\widetilde}
\def\wh{\widehat}
\def\ul{\underline}
\def\nn{\nonumber}
\def\ms{\medskip}
\def\sp{\mbox{Sp}}
\def\re{\mbox{Re}\, }
\def\txs{\textstyle}
\def\dsp{\displaystyle}
\def\ve{V_E({\phi})}
\def\abl{\partial}
\def\vm{V_{M^4}(\phi)}
\def\wew{\left[\frac{l_1^2}{w_1}+\frac{l_2^2}{w_2}\right]}
\def\bacm{(\Phi_{\mu}^2)}
\def\bam{\Phi_{\mu}}
\def\zm{(z_{\mu}^2)}
\def\la{\lambda}
\def\vint{V(M_U^2,M_T^2,M_M^2)}
\def\wewn{\left[\frac{l_1^2}{w_1}+...+\frac{l_n^2}{w_n}\right]}
\def\sleln{{\sum_{l_1,...,l_n=-\infty}^{\infty}}\!\!\!\!\!\!\!^{\prime}}
\def\slel{\sum_{l_1,...,l_n=-\infty}^{\infty}}
\newcommand{\back}{\phi ^{2}}
\newcommand{\bac}{\phi}
\newcommand{\pot}{V(\Phi^2)}
\newcommand{\suz}{\sum_{l_1,l_2=-\infty}^{\infty}}
\newcommand{\sun}{{\sum_{l_1,l_2=-\infty}^{\infty}}\!\!\!\!\!^{\prime}}
\newcommand{\reals}{\mbox{${\rm I\!\!R }$}}
\newcommand{\nats}{\mbox{${\rm I\!\!N }$}}
\newcommand{\intgs}{\mbox{${\rm Z\!\!Z }$}}
\newcommand{\komplex}{\mbox{${\rm I\!\!\!\!C }$}}

\begin{titlepage}

\title{\begin{flushright}
{\normalsize UB-ECM-PF 94/17 }
\end{flushright}
\vspace{2cm}
{\Large \bf On the instability of the vacuum in multidimensional
scalar theories}}

\author{E. Elizalde\thanks{And Center for Advanced Studies, CSIC,
Cam{\ii} de Santa B\`arbara, 17300 Blanes, Spain. E-mail address:
eli@zeta.ecm.ub.es},
K. Kirsten\thanks{Alexander von Humboldt-fellow. E-mail address:
klaus@zeta.ecm.ub.es}  and
Yu.~Kubyshin\thanks{On leave of absence from Nuclear
Physics Institute,
Moscow State University, 117234 Moscow, Russia.}
\thanks{E-mail address: kubyshin@ecm.ub.es} \\
Departament d'ECM, Facultat de F{\ii}sica
\\ Universitat de Barcelona, Av. Diagonal 647, 08028 Barcelona \\
Catalonia, Spain }

\date{October 14, 1994}

\maketitle

\begin{abstract}

The 1-loop effective potential in a scalar theory with quartic
interaction on the space $M^{4} \times T^{n}$ for $n=2$ is
calculated and is shown to be unbounded from below.
This is an indication of a possible instability of the vacuum
of the $\lambda \phi^{4}$ model on $M^{4}$, when it is
regarded as a low energy sector of the theory obtained by
dimensional reduction of the original six-dimensional one.
The issue of stability
for other values of the number $n$ of extra dimensions is also
discussed.

\end{abstract}
\end{titlepage}

\section{Introduction}

It is well known that the quantum properties of  models in
quantum field theory strongly depend on the topology of
the spacetime on which they are formulated and on the boundary
conditions imposed on it. An example of such manifestations is the
influence that non-trivial spacetime topology  has on
the behaviour of the effective potential \cite{veff}, \cite{elikir},
and, in particular, the issue of topological mass
generation \cite{top-mass}, \cite{kir93}. The Casimir effect
\cite{Casimir} is another phenomenon of this type (for a report, see
\cite{plunien}).

In this article we consider
a scalar model with quartic interaction $\lambda \phi^{4}$
defined on $M^{4}\times T^n$, which is regarded as a
laboratory for studying certain quantum properties of field
theories in multidimensional space-times. The latter are viewed
as important candidates for unification
of fundamental interactions (see, for example,
\cite{KK} and \cite{KK-review} for reviews). One of the attractive
features of this approach is that it
provides a natural mechanism for the appearence of scalar Higgs fields
as extra dimensional components of multidimensional gauge fields with
the Higgs potential being a part of the multidimensional
non-abelian gauge action \cite{Manton}. Much work has been done on the
analysis of symmetries and structure of Kaluza-Klein type theories
and on the possibility of obtaining the Standard Model of Grand
Unified
Theories by dimensional reduction (see \cite{dim-red}, \cite{zoupanos}
for reviews). Various issues of quantum properties of multidimensional
theories were studied in \cite{KK-quantum}.

It is known that a multidimensional model can be re-formulated as a
model in four dimensions with an infinite number of fields or modes. The
sector of the zero mode with the lightest mass usually coincides with
the model in four dimensions with the same type of fields and
interactions as
the initial multidimensional one, that is the $\lambda \phi^{4}$ theory
on $M^{4}$ in our case. In this paper we consider the 1-loop effective
potential of the zero mode field calculated in the complete
theory with all modes.
As it is well known by now, the effective
potential is a very important tool in considering stability questions
and the phenomenon of symmetry breaking (see for example \cite{brand}).
We will be interested specifically in the
case $n=2$, though the analysis of the asymptotics of the potential
for large $\phi$ will be carried out for arbitrary $n$. It is obvious,
that since the theory considered for $n \ge 1$ is non-renormalizable,
renormalizations of the mass and the quartic coupling are not enough for
eliminating all ultraviolet divergences. One needs to impose
additional conditions to obtain the renormalized effective potential,
and we will suggest a prescription for doing this at the 1-loop level.
This prescription
will guarantee that in the limit of weak fields the potential of the
complete theory approaches the potential of the corresponding four
dimensional theory, i.e. the Coleman-Weinberg effective potential
\cite{coleman-weinberg}. Namely, when
$\lambda \phi^{2} L /m \ll 1$, where $L$ is the characteristic size of
the torus, these two potentials differ
by higher powers of
$\phi$ only. Thus, the physics described by these two potentials will be
the same in such a limit.
For stronger fields the effects due to the non-trivial
spacetime topology become significant and the
behaviour of the effective potentials turns out to be quite different.
It is worth mentioning here that analogous renormalization prescriptions
for the Green functions imply that physical observables, like for
example the cross sections, of the $\lambda \phi^{4}$ theory on $M^{4}$
coincide with those of the corresponding multidimensional model in the
low-energy limit \cite{dkp}.

To see the effect that the extra dimensions with non-trivial topology
have on the effective potential, we restrict our study here to a simple
scalar model, calculate
the effective potential of the zero mode in the 1-loop approximation
and compare it with the known result of Coleman and Weinberg for the
usual four-dimensional $\lambda \phi^{4}$ model \cite{coleman-weinberg}.
The model might be relevant for the
description of the Higgs sector of some Kaluza-Klein type extensions of
the Standard Model and in spite of its simplicity it features
some important properties of Kaluza-Klein models.

We will do our calculations for the case of periodic and antiperiodic
boundary conditions for the scalar field on $T^{2}$ and will also
analyze the
mixed case of the field periodic in one dimension and antiperiodic in
the other,
taking into account
also interactions between the different types of scalar fields.

We adopt the point of view that the size $L$ of the space
of extra dimensions (the two-dimensional torus in our case) can
take values in a wide range, instead of choosing it of the order of the
Planck length, as it is suggested by most of spontaneous
compactification solutions.
The estimates, carried out in \cite{kolb}, give the bound on the
present value $L^{-1} < 10^{6}$ GeV.
Other motivations are related to results of papers
\cite{kapl-ant}
and suggest that $L^{-1}$ should be of the order of the supersymmetry
breaking scale $M_{SUSY}$.
Thus, for a class of string
models with orbifold compactification the bounds on $L$
following from high-energy experiments give $L^{-1} \sim 1
\mbox{TeV}$ \cite{antoni}. The parameter
$L$ in our analysis plays a role similar to that of the
inverse temperature $T^{-1}$ in models of non-zero temperature.
We would like to mention that
calculations of the effective potential in theories with non-zero
temperature, when the spacetime can be effectively taken as $M^{3}
\times S^{1}$ and $L = T^{-1}$ is the radius of the circle $S^{1}$, can
be found in \cite{temperature}.

One important remark is in order. In our calculations gravity is
considered classically
as a non-trivial spacetime background. This makes sense if we limit
ourselves to the scalar field sector only and assume that the geometry
of the space of extra dimensions and its size $L$ are classical
variables and are determined by some
additional mechanism. Results of ref.~\cite{appel} show that in a
more
general theory including gravity, where $L$ is considered as a dynamical
variable, contributions from gravitons are significant if $L$ is
of the order of the Planck length. Then the
total effective potential should include the contribution from scalars,
we calculate here, and the graviton part,
calculated in \cite{appel}, and can be used
for the analysis of the spontaneous compactification solutions of the
theory
of gravity coupled to scalars. To be relevant to the cosmology of the
early Universe the effective potential should be further modified to
include contributions due to non-zero temperature. The finite
temperature graviton effective potential on the spacetime $M^{4} \times
T^{n}$ was calculated by Appelquist et al.~\cite{ACM-83} and temperature
corrections were found to be quite important (see also \cite{RR-83}).
The finite temperature effective potential for the free scalar theory
on the same spacetime was studied in \cite{RR-veff} and the stability
of the total effective potential was analyzed in
\cite{torus-stability}.
The calculation of temperature corrections to the scalar effective
potential
in the model with interaction which is under consideration in this
article and the analysis of
spontaneous compactification solutions and their stability for that case
would be of great interest.

The paper is organized as follows. In Sect.~2 we describe the model
and discuss its unrenormalized effective potential.
In Sect.~3 the renormalized effective potential
on
the spacetime $M^{4} \times T^{2}$ with periodic boundary
conditions for the scalar field is calculated using zeta-function
techniques and its properties for weak and strong fields are studied.
Representations for the zeta-functions used for these calculations are
given in Appendix A.
Sect.~4 is devoted to an
extension of our results to the case of antiperiodic and mixed boundary
conditions for the scalar field including interactions between the
different types of scalar fields. The asymptotics of the
potential for large
$\phi$ and its boundedness or unboundedness from below for
more general spaces $M^{4} \times T^{n}$ are analyzed  in Sect.~5.
Finally, Sect.~6 contains the discussion of the results.
Some auxilliary results on the Epstein-type zeta-functions and
formulas of the 1-loop contributions for the renormalized effective
potential are presented in Appendices A-C.

\section{Description of the model and the effective potential}

Let us consider a one component scalar field on the
$6$-dimensional
manifold $E=M^{4}\times T^{2}$, where $M^{4}$ is Minkowski
space-time and $T^{2}$ is the two-dimensional torus
with circumferences $L_1$ and $L_2$. In spite of its
simplicity this model captures some interesting features of
both classical and quantum properties of
multidimensional theories. The classical action is given by
\begin{equation}
S= \int_{E} d^{4}x d^{2}y \left[\frac{1}{2}\left(\frac{\p \hat{ \f}
(x,y)}{\p x^{\mu}}\right)^2+\frac{1}{2} \left( \frac{\p \hat{ \f} (x,y)}
{\p y^{i} } \right)^2 -
 \frac{1}{2} m_{0}^{2} \hat{ \f} ^{2}(x,y) -
\frac{\hat{\lambda}}{4!} \hat{ \f} ^{4} (x,y) \right],
\label{eq:action0}
\end{equation}
where $x^{\mu}, \mu = 0,1,2,3,$ are the coordinates on
$M^{4}$ and $y^{1}$ and $y^{2}$ are the
coordinates on $T^{2}$, $0
\leq y^{1} <  L_{1}$, $0 \leq y^{2} < L_{2}$.
To re-interpret this model in four-dimensional terms we
make the Fourier expansion of the field $\hat{ \f} (x,y)$,
\begin{equation}
 \hat{ \f} (x,y) = \sum_{N} \phi _{N} (x) Y_{N} (y),
\label{eq:expansion}
\end{equation}
where $N=(l_{1},l_{2})$, $-\infty < l_{i} < \infty$ and
$Y_{N}(y)$ are the
eigenfunctions of the Laplace operator on the internal space
\begin{equation}
    Y_{(l_{1},l_{2})}  =  \frac{1}{ \sqrt{L_{1} L_{2}}}
    \exp \left[  2\pi i
\left(\frac{l_{1}y^{1}}{L_1}+\frac{l_{2}y^{2}}{L_2}
     \right) \right].
\label{eq:laplace}
\end{equation}
Substituting this expansion
into the action and integrating over $y$, one obtains
\begin{eqnarray}
   S & = & \int_{M^{4}} d^{4} x \left\{ \frac{1}{2} \left( \frac{\p
\f _{(0,0)} (x)} {\p x^{\mu}} \right)^{2}
- \frac{1}{2} m_{0}^{2} \f _{(0,0)}^{2}(x) -
\frac{\lambda}{4!} \f _{(0,0)}^{4} (x) \right.  \nonumber  \\
     &  & + \sum_{N>0} \left[ \frac{\p \f_{N}^{*} (x)}{\p
x^{\mu}}
\frac{\p \f_{N} (x)}{\p x_{\mu}} - M_{N}^{2} \f _{N}^{*} (x)
\f _{N} (x) \right]  \nonumber  \\
     &  & - \left. \frac{\lambda}{2}
\f _{(0,0)}^{2} (x) \sum_{N > 0} \f_{N}^{*}(x)
\f_{N}(x) \right\} - S'_{int},
\label{eq:action1}
\end{eqnarray}
where the four-dimensional coupling constant $\lambda$ is
related to the
multidimensional one $\hat{\lambda}$ by
$ \lambda = \hat{\lambda} / {\mbox volume} (T^{2})$.
In eq. (\ref{eq:action1})
$S_{int}'$ contains all terms of the third and fourth
powers in
$\f_{N}$ with $N^{2} > 0$. We see that the model
includes one
real scalar field $\f_{(0,0)}$ describing a
light particle of mass $m_{0}$, and an infinite set
(``tower") of massive complex fields $\f_{N}(x)$ corresponding
to heavy particles, or pyrgons, of mass
given by
\begin{equation}
  M_{N}^{2} = m_{0}^{2} + l_{1}^{2}M_{1}^2 + l_{2}^{2}M_{2}^2 ,
  \label{eq:mass}
\end{equation}
where $M_{i} = 2\pi /L_{i}$.

In the spacetime considered here one may assume a constant classical
background field $\hat{\phi}_{cl}$, and the quantum fluctuations
$\varphi = \hat{\phi} -\hat{\phi}_{cl}$ around the background field
in the linear approximation satisfy the equation
\beq
(-\Delta +\Phi^2) \varphi =0,\label{l1}
\eeq
with the effective mass $\Phi^2=m_0^2+
(1/2)\hat{\lambda}\hat{\phi}_{cl}^2=m_0^2+(1/2)\lambda\back$,
where we have expressed the effective mass through the four dimensional
quantities. The
effective potential for the field $\bac$
including one-loop quantum effects is then given by the function
\beq
V_E(\bac )=\frac{1}{2} m_{B}^2\back +\frac 1 {4!} \lambda_{B}\bac
^4+V_2\pot,         \label{l2}
\eeq
where $m_{B}$ and $\lambda_{B}$ are the bare mass and coupling constant
and the quantum corrections amount to
\beq
(V_4V_2)\pot =\frac 1 2 \ln\det \left(\frac{-\Delta
+\Phi^2}{\kappa^2}\right),   \label{l3}
\eeq
which is the functional determinant arising from the
integration over the quantum fluctuations, $V_4$ and $V_2$ are
the volumes of $M^4$ and $T^2$ respectively and $\kappa$
is a scale which arises in this formalism. General methods for
the calculation of the effective potential were developed in
\cite{coleman-weinberg}, \cite{wein}; for calculations of the effective
potential in theories in spaces with nontrivial topologies
and for recent
results in the Standard Model at non-zero temperature see \cite{actor}
and \cite{arnold}, respectively.

For the calculation of the functional determinant, eq.~(\ref{l3}), we
will use the zeta-function prescription
\cite{hawking}. In this scheme
\beq
V_2\pot =-\frac 1 2 \left[\zeta (0;M^{4}\times T^2)
\ln\kappa^2+\zeta'(0;M^{4}\times T^2)\right], \label{l4}
\eeq
where the prime denotes
differentiation with respect to $s$, see eq.~(\ref{zeta}).
$\zeta (s;M^{4}\times T^2)$ is
the zeta function associated with the operator (\ref{l1}) with periodic
boundary conditions for the field $\varphi$. This means that
for $\Re e\,\, s >3$,
\beq
\zeta(s;M^{4}\times T^2) =\frac{1}{(2\pi)^{4}} \suz \int
d^4k\,\,\left[\left(\frac{2\pi l_1}{L_1}\right)^2+
\left(\frac{2\pi l_2}{L_2} \right)^2+
 k^2+\Phi^2\right]^{-s}   \label{zeta-gen}
\eeq
or, performing the $k$-integration,
\beq
\zeta (s;M^{4}\times T^2)=\frac{\pi^2}{L^4}\left(\frac L
{2\pi}\right)^{2s}\frac{\Gamma(s-2)}{\Gamma(s)}Z_2^{v^2}(s-2;w_1,w_2),
\label{zeta}
\eeq
where we have introduced the dimensionless parameters $v^2=(L\Phi /2\pi
)^2$ and $w_i=(L/L_i)^2$ ($L$ is an arbitrary length parameter). The
generalized Epstein zeta-function $Z_2^{v^2}(\nu; w_1,w_2)$ has
the representation
\beq
Z_2^{v^2}(\nu; w_1,w_2)=\suz [w_1l_1^2+w_2l_2^2+v^2]^{-\nu},
\label{l6}
\eeq
valid for $\Re e\,\,\nu >1$, and is very well
studied by now \cite{elikir,kir93}. We have
listed
its essential properties in Appendix A, indicating very briefly
their derivation. Comparing with eq. (\ref{a1}), we use here the
notation $Z_2^{v^2}(\nu; w_1,w_2) = Z_2^{v^2}(\nu; w_1,w_2 ; 0,0)$.
{}From (\ref{l4}), the quantum potential is found to be
\begin{eqnarray}
V_2\pot &=&-\frac 1 4
\frac{\pi^2}{L^4}\left\{{Z_2'}^{v^2}(-2;w_1,w_2)-\frac{\pi} 3
v^6\frac{L_1L_2}{L^2}\left[2\ln
\left(\frac{L\kappa}{2\pi}\right)+\frac 3 2 \right]\right\}\nn\\
&=&\frac{\pi^3L_1L_2}{12L^6}v^6\left[\frac{11} 6
-\ln\left(\frac{\Phi} \kappa\right)^2\right]   \label{l7}\\
& &-\frac{L_1L_2}{L^6}v^3\sun\hspace{0.2cm}\left[\frac{l_1^2}{w_1}+
\frac{l_2^2}{w_2}\right]^{-\frac 3 2}K_3\left(2\pi v
\left[\frac{l_1^2}{w_1}+\frac{l_2^2}{w_2}\right]^{\frac 1 2}\right),
\nn
\end{eqnarray}
where in the last equality eq.~(\ref{a4}) has been used.
Eq.~(\ref{l7}) re\-pre\-sents the non-re\-nor\-ma\-liz\-ed one-loop
contribution to the effective potential of the theory. As one can
easily check, in spite of the presence of the parameter $L$ in eq.
(\ref{l7}), the potential  does not depend on it actually.
The dependence on $\kappa$ will disappear after
renormalization.

\section{Renormalization of the one-loop effective potential}

In the previous section we have derived the regularized one-loop
contribution to the
effective potential $\pot$ for a scalar theory defined on the manifold
$E=M^4\times T^2$. In accordance with the discussion in the
Introduction, we need to
consider a potential of the following general form
\bea
\ve &=&\frac 1 2 m_0^2\back +\frac 1 {4!} \lambda\bac ^4\label{r1}\\
& &+A+\frac 1 2 B \back +\frac 1 {4!} C \bac ^4 +\frac 1 {6!}
D \bac ^6 +V_2\pot , \nn
\eea
where $m_{0}$ and $\lambda$ are the renormalized mass and coupling
constant, respectively, and the bare parameters are understood
to be defined as $m_B^2 =m_0^2+B$, $\lambda_{B}=\lambda + C$.
The constant $D$ subtracts the additional divergences arising due
to the six-dimensional nature of the theory. For the $\lambda \phi^{4}$-
model on $M^{4}$ to be a low-energy limit of our
multidimensional theory, we impose the following
renormalization conditions:
\bea
\ve |_{\bac =0}&=&0=\vm|_{\bac =0},   \label{r21} \\
\frac{\abl ^2\ve}{\abl \bac^2}\left|_{\bac =0}\right.&=&m_0^2 =
\frac{\abl ^2\vm}{\abl \bac^2}\left|_{\bac =0}\right., \label{r22}\\
\frac{\abl ^4\ve}{\abl \bac^4}\left|_{\bac =\mu}\right.&=&\lambda =
\frac{\abl ^4\vm}{\abl \bac^4}\left|_{\bac =\mu}\right.,  \label{r23} \\
\frac{\abl ^6\ve}{\abl \bac^6}\left|_{\bac =\mu}\right.&=&
\frac{\abl ^6\vm}{\abl \bac^6}\left|_{\bac =\mu}+h\right. , \label{r24}
\eea
where, for the sake of generality, we have introduced in the last
equation the additional
coupling of the $\phi^{6}$-vertex, that appears in principle
because of the
non-renormalizability of the theory.
$V_{M^4}(\bac )$ is the Coleman-Weinberg potential in the four
dimensional Minkowski spacetime before renormalization. It reads
\cite{coleman-weinberg}
\beq
V_{M^4}(\bac ) =\frac 1 2 m_0^2 \bac ^2+ \frac 1 {4!} \lambda \bac^4 +
\frac 1 {64\pi^2}\Phi^4\left[\ln\left(\frac{\Phi}{\kappa}\right)^2-
\frac 3 2\right]+A'+\frac 1 2 B'\bac ^2+\frac 1 {4!} C'\bac ^4,
\label{r3}
\eeq
with $A',B',C'$ being determined by the conditions ~(\ref{r21})-
(\ref{r23}). For zero mass, $m_0^2=0$,
one has to choose $\mu \neq 0$.
As we will see in what follows, it is useful to consider the function
\bea
H(\Phi^2; u_1,u_2 )&=&-\frac{\Phi^3L_1L_2}{8\pi^3 L^3}\sun
\exp\{2\pi i[l_1u_1+l_2u_2]\}\label{r4}\\
& &\qquad\qquad\wew ^{-\frac 3 2}K_3 \left(L\Phi\wew^{\frac 1
2}\right),\nn
\eea
where the phase factor has been introduced with regard to the different
boundary conditions treated later on. For the moment we only need
$H(\Phi^2):=H(\Phi^2;0,0)$.

Denoting by $H^{(n)}(\Phi^2)$ the $n$-th derivative with respect to
$\Phi^2$, eqs.~(\ref{r21})-(\ref{r24}) determine the
counterterms to be
\begin{eqnarray}
A(m_0^2)&=&-H(m_0^2)-\frac{L_1L_2}{768\pi^3}m_0^6\left[\frac{11}
6-\ln\left(\frac{m_0} \kappa\right)^2\right],
                                             \label{r5} \\
B(m_0^2)&=&-\la H^{(1)}(m_0^2)-\frac{\la L_1L_2}{768\pi^3}3m_0^4
\left[\frac
3 2 -\ln \left(\frac{m_0} \kappa\right)^2\right],
                                              \label{r6}\\
C&=&C^{(1)}(m_0^2,\la)\nn\\
& &-\frac{\mu^2}{32\pi^2}\left\{\frac{15} 2 \frac{\la ^3}{\bam^2}-\frac
{45} 2 \frac{\la ^4\mu^2}{\bam^4}-\frac{15 \la ^5\mu^4}{\bam^6} -3
\frac{\la^6\mu ^6}{\bam^8}\right\}\nn\\
& &-\frac 1 2 h\mu^2,                           \label{r7}\\
D&=&D^{(1)}(m_0^2,\la )
+\frac {15\la^3}{32\pi^2\bam^2}
-\frac{45\la^4\mu^2}{32\pi^2\bam^4} \nn\\
& &+\frac{30\la^5\mu^4}
{32\pi^2\bam^6}
-\frac{6\la^6\mu^6}{32\pi^2\bam^8}
+h,                                              \label{r8}
\end{eqnarray}
with $\bam^2=m_0^2+\frac 1 2 \la \mu^2$.
Here $C^{(1)}(m_0^2,\la)$ and $D^{(1)}(m_0^2,\la )$ denote the
contribution of the one-loop quantum correction $V_2V(\Phi^2 )$ to the
counterterms. Explicitly they read
\begin{eqnarray}
C^{(1)}(m_0^2,\la )&=&
-3\la ^2 H^{(2)}\bacm +\frac 3 2 \la ^3\mu^2 H^{(3)}\bacm +\frac{43} 2
\la
^4\mu^4 H^{(4)}\bacm      \nn\\
& &+\frac{15}
2 \la ^5 \mu^6 H^{(5)} \bacm + \frac 1 2 \la^6 \mu^8
H^{(6)}\bacm               \label{onec}\\
& &
+\frac{L_1L_2}{768\pi^3}\left\{18\la ^2 m_0^2\ln \left(\frac{\bam}
{\kappa}\right)^2-129\frac{\la ^4
\mu^4}{\bam^2}-18\la ^2\bam ^2+\frac{45\la ^5\mu^6}{\bam^4}-\frac{6\la
^6\mu^8}{\bam^6}\right\} ,       \nn\\
D^{(1)}(m_0^2,\la )&=&
-15 \la ^3H^{(3)}\bacm -45 \la^4\mu^2 H^{(4)}\bacm -15 \la^5\mu^4
H^{(5)}\bacm -\la ^6 \mu^6 H^{(6)}\bacm           \nn\\
& &  \hspace{-1cm}+\frac{90\la ^3L_1L_2}{768
\pi^3}\ln\left(\frac{\bam}\kappa\right)^2
+\frac{270\la^4\mu^2L_1L_2}{768\pi^3\bam^2}
-\frac{90\la^5\mu^4L_1L_2}{768\pi^3\bam^4}
+\frac{12\la^6\mu^6L_1L_2}{768\pi^3\bam^6}.
                                                \label{oned}
\end{eqnarray}
Use of eqs.~(\ref{r5})-(\ref{r8}) in eq.~(\ref{r1}) yields the
renormalized effective potential. As we will see, the dependence on the
arbitrary scale $\kappa$ has vanished.

For the presentation of the effective potential in the case
$m_{0} \neq 0$ it is convenient to introduce dimensionless
parameters. As such, we choose
\beq
x=\frac{\la \back}{2m_0^2};\,\,\,y=\frac{\la \mu^2}{2m_0^2};\,\,\,
g=\frac{\la}{64\pi^2};\,\,\,\nu =m_0^2L_1L_2;\,\,\,\xi
=\frac{hm_0^2}{90\la^2}.
\nn
\eeq
Furthermore let us rewrite the terms containing the McDonald (or
modified Bessel) functions in
the form
\beq
H(\Phi^2 ) =\frac{L_1L_2}{64\pi^2L^6}E(L^2 \Phi^2)  \nn
\eeq
and, finally, let us introduce the following dimensionless function
characterizing the effective potential:
\[
v_{E}(x)  =\frac{\la \ve}{m_0^4}.
\]
We find that
\beq
v_{E}(x) =v^{(0)}(x)+v^{(1)}(x),     \label{r9}
\eeq
where
\bea
v^{(0)}(x)
&=&x+\frac 1 6 x^2
-\frac 2 3 gyx^2\left[\frac{15} 2 \beta-45y\beta^2+60
y^2\beta^3-24y^3\beta^4\right]-15 y\xi x^2+x^3\xi \nn\\
& &+\frac 1 {45} gx^3\left[15\beta-90 y\beta^2 +120 y^2\beta^3 -48
y^3\beta^4\right]                        \label{nr9}
\eea
and
\bea
\lefteqn{ v^{(1)}(x)=\frac 1
{12\pi} g\nu\ln\left(\frac{m_0}{\Phi}\right)^2+\frac 1 {12\pi}
gx\nu\left[1+3\ln\left(\frac{m_0}{\Phi}\right)^2\right]    }\nn\\
& &+\frac{\nu}
{72\pi}gx^2\left[18\ln\left(\frac{\bam}{\Phi}\right)^2+15
-18y-516y^2\beta +360 y^3\beta^2
-96 y^4\beta^3\right]\nn\\
& &+\frac 1 {1080\pi} g\nu x^3\left[ 90
\ln\left(\frac{\bam}{\Phi}\right)^2+165+ 540y\beta-360 y^2\beta^2 +96
y^3\beta^3\right]                         \label{onep}\\
& &+\left(\frac{L_1L_2}{L^2}\right)\left\{-\frac g {\alpha^4}
E(m_0^2L^2)
-\frac{gx}{\alpha^2}E^{(1)}(m_0^2L^2) +\frac
g{\alpha^4}E(z^2)\right.\nn\\
& &+\frac 1 6 gx^2\left[-3E^{(2)}\zm +3y \alpha^2E^{(3)}\zm +86
y^2\alpha^4 E^{(4)}\zm \right.\nn\\
& &\left.\qquad\qquad +60 y^3\alpha^6 E^{(5)}\zm +8y^4 \alpha^8
E^{(6)}\zm \right]\nn\\
& &\left.-\frac g {90} x^3 \alpha^2\left[ 15 E^{(3)}\zm +90 y \alpha^2
E^{(4)} \zm +60 y^2 \alpha^4 E^{(5)}\zm +8 y^3 \alpha^6
E^{(6)}\zm\right]\right\},  \label{v1}
\eea
where we use the notations
\[
\beta=\frac 1 {1+y};\,\,\, \alpha =m_0 L = \sqrt{\nu w_{1}w_{2}};
\,\,\,
z=L\Phi=\alpha (1+x); \,\,\,z_{\mu}=L\Phi_{\mu}=\alpha (1+y).
\]
The reason for the separation chosen in eq. (\ref{r9})
will become clear from the next section.

Let us now study the expression for the renormalized
effective potential with $m_{0} \neq 0$.
First we analyze the asymptotics of eq. (\ref{v1}) for large $x$. Here
we restrict ourselves to the case $\mu = 0$ and
$L_{1}=L_{2}=L$. The leading terms are:
\beq
   v_{E} \approx - \frac{g \nu x^{3}}{12 \pi} \ln (cx)
   + \frac{g+3\xi}{3} x^{3}, \label{asymp}
\eeq
where $c$ is given by: $\ln c = -11/6 +
E^{(3)}(\nu)/6$. We see that the potential is unbounded
from below for large values of the field and that such
behaviour is a consequence of the six-dimensional
nature of the model. Indeed, the leading asymptotics
$v_{E} \sim - g \nu x^{3} \ln x /12 \pi$ is given by the
first term in eq. (\ref{l7}).
Since the renormalization conditions (\ref{r21})-(\ref{r24})
amount to
changing terms at most $ \sim x^{3}$, the asymptotics are not affected
by renormalization.

As mentioned above, the representation eqs.~(\ref{r9})-(\ref{v1}) is
useful
for large fields and finite $\nu$. However, in order to see the limit
of small fields and/or how the Coleman-Weinberg potential results
when extra dimensions disappear in the limit $L \rightarrow 0$, it
is more convenient to use another representation. To derive it, notice
that the zeta function associated with the six-dimensional operator
(\ref{l1}), can be decomposed as follows:
\beq
  \zeta(s;M^{4}\times T^2) = \zeta(s;M^{4}) + \tilde{\zeta}(s;
  M^{4} \times T^{2}).    \label{repr2}
\eeq
Here $\zeta(s;M^{4})$ is the 1-loop integral on
$M^{4}$ given by the term with $l_{1}=l_{2}=0$ in
eq. (\ref{zeta-gen}). After
performing the $k$-integration it is equal to
\beq
\zeta(s;M^{4}) = \frac 1 {16 \pi^2 (1-s)(2-s)} (\Phi^{2})^
{2-s}.   \nn
\eeq
$\tilde{\zeta}(s;M^{4}\times T^{2})$ is the rest of the sum (\ref{zeta-gen})
and contains contributions due to the extra dimensions.

Maintaining the decomposition corresponding to (\ref{repr2}) at
each step,
we perform the renormalization as before and get the result
(for $m_{0} \neq 0$):
\beq
  v_{E}(x) = v_{M^{4}}(x) + \xi x^{3} - 15 \xi y x^{2} +
  \Delta v_{1}(x) + \Delta v_{2}(x) + \Delta v_{3}(x).
                                   \label{vE-small}
\eeq
Here
\bea
  v_{M^{4}}(x) & = & x + \frac{x^{2}}{6} - gx -
             \frac{3}{2} g x^{2} + g(1+x)^{2} \ln(1+x)  \nn \\
             & & - g x^{2} \ln(1+y) - \frac{4}{3} g x^{2} y
                 \frac{3+2y}{(1+y)^{2}}   \label{vM4-small}
\eea
is the part corresponding to the re\-nor\-ma\-li\-zed one-loop
Co\-le\-man-Wein\-berg
po\-ten\-ti\-al $V_{M^{4}}(\phi)$ (with non-zero mass) through the relation
$v_{M^{4}}(\lambda \phi^{2}/(2 m_{0}^{2})) = (\lambda / M_{0}^{4})
V_{M^{4}}(\phi)$. The term $\Delta v_{1}(x)$ is equal to
\bea
\Delta v_{1}(x) & = & g \sun \hspace{2mm}
       \left[ \frac{1}{\nu^{2}} (l_{1}^2 w_{1} + l_{2}^{2} w_{2}
       + \nu (1+x))^{2} \ln \frac{l_{1}^2 w_{1} + l_{2}^{2} w_{2}
       + \nu (1+x)}{l_{1}^2 w_{1} + l_{2}^{2} w_{2} + \nu }
       \right.                                          \nn  \\
       & - & \left. \frac{x}{\nu} (l_{1}^2 w_{1} + l_{2}^{2} w_{2}
       + \nu) - \frac{3}{2} x^{2} - \frac{1}{3} \frac{x^{3} \nu}
       {(l_{1}^2 w_{1} + l_{2}^{2} w_{2} + \nu)} \right].
                                               \label{dv1-small}
\eea
The terms $\Delta v_{2}(x)$ and $\Delta v_{3}(x)$ are
given in Appendix C. As one can see they vanish when $y=0$.
The analogous representation
for the case $m_{0} = 0$ is presented in Appendix C too.
Using these formulas we obtain the behaviour of the potential
for small $x$ and small $\nu$.

We first consider the case $m_0\neq 0$ in more detail and for the
sake of simplicity choose $\mu =0$.  Then $\Delta v_{2} = \Delta
v_{3} =
0$. For $x \nu \ll 1$ \[ \Delta v_{1} = - \frac{g}{12} \alpha^{4} x^{4}
\sun \hspace{3mm} \frac{1}{(w_{1} l_{1}^{2} + w_{2} l_{2}^{2} +
\nu)^{2}} + \frac{1}{\alpha^{4}}{\cal O}(\nu^{5} x^{5})     \]
and we obtain
\beq
v_{E} = v_{M^{4}}(x) + \xi x^{3}
- \frac{g}{12} \alpha^{4} x^{4} Z_{2}^{\alpha^{2}}(2;w_{1},w_{2}) +
{\cal O}(x^{5}).        \label{vE-expan}
\eeq

{}From the formula above we see that, due to the renormalization
conditions imposed, the effective potential $V_{E}$ gives the
values of $m_{0}$ and $\lambda$ at the subtraction point
$\phi_{0}=0$. Moreover, these conditions also assure that
the contributions of the heavy modes are of the order
$\phi_{0}^{8}$ for small fields, namely when
$x \nu \ll 1$ or $\phi_{0}^{2} \ll 2/(\lambda L^{2})$.

On the other hand, from eq. (\ref{vE-expan}) we see that, for any
fixed $x$ and $\nu \ll 1$ (we assume that $w_{1}$
and $w_{2}$ are fixed), we get
\[
v_{E}(x) = v_{M^{4}} + \xi x^{3} - \frac{g \nu ^{2} w_{1}^{2}
w_{2}^{2} x^{4}}{12} Z_2(2;w_1,w_2) + {\cal O}(\nu ^{3}),
\]
where
\[ Z_2(s;w_1,w_2) = \sun \hspace{3mm}
\frac{1}{(w_{1}l_{1}^{2}+w_{2}l_{2}^{2})^{s}}.   \]
It is clear that for $\xi =0$ the four-dimensional Coleman-Weinberg
potential is recovered in the limit $L \rightarrow 0$, i.e. when the
space of extra dimensions shrinks down to zero.

In the massless case, $m_0^2=0$,
using the expansion of the Epstein-type zeta function for small values
of the field, see Appendix A, eq.~(\ref{a7}), we find
\bea
\lefteqn{
V_2 V(\Phi^2 ) =-\frac 1 2 \frac{\pi ^2}{L^4}\times   }  \nn \\
& &\left\{-\frac{1}{2} v^4 \ln v^2 +\frac{1}{2} {Z_{2}'}
(-2;w_1,w_2) +v^2
{Z_2'} (-1;w_1,w_2)\right.\nn\\
& &+\frac 1 2 v^4 \left[\frac 3 2
+{Z_2'} (0;w_1,w_2)\right]\label{vier1}\nn\\
& &\left.-\frac 1 6 v^6 \left[  PP\,\,Z_2(1;w_1,w_2)+\frac{2\pi
L_1L_2}{L^2}
\ln\left(\frac{L\kappa}{2\pi}\right)\right]+g(v)\right\}, \nn
\eea
with
\beq
g(v)=\sum_{j=2}^{\infty}(-1)^j\frac{(j-1)!}{(j+2)!}v^{2j+4}
Z_2^{v^2}(j;w_1,w_2).                  \label{vier2}
\eeq
Here, $PP\,\,Z_2(1;w_1,w_2)$ denotes the finite part of $Z_2
(1;w_1,w_2)$.
Without presenting once more the detailed calculations, by imposing the
renormalization conditions, eqs.~(\ref{r21})-(\ref{r24}), and using this
time the above
representation, the final result for the renormalized effective
potential reads
\bea
\lefteqn{
  V_E(\bac )= V_{M^{4}}(\bac) }\nn\\
& &+\frac 1 {4!} \bac^4\left\{-\frac 1 2 h \mu^2
+\frac{\lambda^2}{128\pi^2}g^{(4)}(v_{\mu})-\frac{\lambda^3L^2\mu^2}{2048
\pi^4}g^{(6)}(v_{\mu})\right\}                \label{vier3}\\
& &+\frac 1 {6!} \bac^6 \left\{h+\frac{ \lambda ^3L^2}{1024 \pi^4}
g^{(6)}(v_{\mu})\right\} -\frac 1 2 \frac{\pi^2}{L^4}g(v).    \nn
\eea
The first line is exactly the Coleman-Weinberg result
\cite{coleman-weinberg}
\beq
V_{M^{4}}(\bac) = \frac 1 {4!} \lambda \bac ^4 +\frac{\lambda^2}{256\pi^2}
\bac ^4 \left[ \ln \left(\frac{\bac}{\mu}\right)^2-\frac{25}
{6}  \right],     \label{vcw}
\eeq
the following terms are corrections to it, the last
one being of the order
${\cal O}((L\phi)^8)$.

For the case $m_{0}=0$ the representation analogous to
eqs. (\ref{vE-small})-(\ref{dv1-small}),
useful for the analysis at small values of the
field or small values of $t=(L \mu)^2$, is the following:
\beq
u_{E}(z) = \frac{4!}{\lambda \mu^4} V_{E}(\bac) = u_{M^4}(z) + \eta
    z^3 -15\eta z^2 + \Delta u_{1}(z) + \Delta u_{2}(z) +
    \Delta u_{3}(z),     \label{uE-small}
\eeq
where $z=\back/\mu^2$, $\eta = h \mu^2 /(30 \lambda)$ and
\[
u_{M^4}(z) = \frac{4!}{\lambda \mu^4} V_{M^4}(\bac)
\]
is again the term corresponding to the Coleman-Weinberg potential
(\ref{vcw}). The terms $\Delta u_{i}(z)$, $i=1,2,3$, are given in
Appendix C.

Using the formulas of this section we have computed the behaviour
of the functions characterizing the effective potential with $w_{1}=
w_{2}=1$.
In Fig. ~1 the plots of $v_{M^4}(x)$ and $v_{E}$ for
three values of $\nu$ and for $y=0$, $\xi=0$ are given.
For small $x$ the curves are very close, whereas for
large $x$ the asymptotic behaviour (\ref{asymp}) is recovered.
It is seen that for smaller values of $\nu$ the function
$v_{E}(x)$ approaches the Coleman-Weinberg potential
$v_{M^{4}}(x)$ for a wider range of $x$.

The plots of the functions $u_{M^4}(z)$ and $u_{E}(z)$ for $t=0.007$,
$0.01$ and $0.015$
are presented in Figs.~2 and 3.
The curves are close to each other for small values of $z$
showing non-trivial minima, whereas with increasing $z$
the functions $u_{E}(z)$ deviate more and more from $u_{M^4}(z)$
and are unbounded from below for large $z$.

Varying $h$ and $w_{1}$, $w_2$ does not change the
behaviour of $v_{E}(x)$ or $u_{E}(z)$ qualitatively.
Essentially this only changes the values of the field
where these functions begin to decrease.

The results of this section show, that the vacuum state $\phi =0$ of the
four dimensional theory might be unstable due to the presence of the
two extra dimensions of the original space-time.

\section{Antiperiodic and mixed boundary conditions on the torus}

Up to now we have only considered periodic boundary conditions
for the scalar field in both directions of the torus $T^{2}$,
see eqs.~(\ref{eq:expansion}), (\ref{eq:laplace}). However,
there are two further
different types of scalar fields: antiperiodic in both directions,
$\hat{\f}(x,y^{1}+mL_{1},y^{2}+nL_{2}) = (-1)^{m+n}
\hat{\f}(x,y^{1},y^{2})$ and
periodic in one direction and antiperiodic in the other
$\hat{\f}(x,y^{1}+mL_{1},y^{2}+nL_{2}) = (-1)^{m}
\hat{\f}(x,y^{1},y^{2})$. We study such possibilities in this section.

In considering them, we restrict ourselves to the case
$L_1=L_2=L$, thus $w_i=1$. This kind of
boundary conditions leads to the general type of Epstein zeta-functions
\beq
Z_2^{v^2}(\nu; w_1,w_2;u_1,u_2)=\suz \left[w_1(l_1-u_1)^2+
w_2(l_2-u_2)^2+v^2\right]^{-\nu}.\label{z1}
\eeq
The relevant properties are summarized in Appendix A. For periodic
boundary conditions we have $u_1=u_2=0$, for antiperiodic boundary
conditions $u_1=u_2=1/2$, and, finally, for the mixed ones $u_1=0$ and
$u_2=1/2$.

For antiperiodic and mixed boundary conditions alone, the analysis
of the previous section makes no sense for the simple reason,
that there is no
constant classical background field with this boundary conditions apart
from $\bac =0$. However, the influence might be present if
one includes interactions between the
different
types of scalar fields \cite{veff-review}. Thus, as a generalization
to eq.~(\ref{eq:action0}), we consider the action
\bea
S & = & \int_{E} d^{4}x d^{2}y \left[\frac{1}{2}(\frac{\p \hat{\f} _U
(x,y)}{\p x^{\mu}})^2+\frac{1}{2} \left( \frac{\p \hat{\f} _U(x,y)}
{\p y^{i} } \right)^2 -
 \frac{1}{2} m_{U}^{2} \hat{\f}_U ^{2}(x,y) -
\frac{\hat{\lambda}_U}{4!} \hat{\f}_U ^{4} (x,y) \right.\nn\\
& &+\frac{1}{2}(\frac{\p \hat{\f} _T
(x,y)}{\p x^{\mu}})^2+\frac{1}{2} \left( \frac{\p \hat{\f} _T(x,y)}
{\p y^{i} } \right)^2 -
 \frac{1}{2} m_{T}^{2} \hat{\f}_T ^{2}(x,y) -
\frac{\hat{\lambda}_T}{4!} \hat{\f}_T ^{4} (x,y) \nn\\
& &+\frac{1}{2}(\frac{\p \hat{\f} _M
(x,y)}{\p x^{\mu}})^2+\frac{1}{2} \left( \frac{\p \hat{\f} _M(x,y)}
{\p y^{i} } \right)^2 -
 \frac{1}{2} m_{M}^{2} \hat{\f}_M ^{2}(x,y) -
\frac{\hat{\lambda}_M}{4!} \hat{\f}_M ^{4} (x,y)
                                         \label{z2}\\
& &\left.-\frac 1 4 \hat g _{TU}\hat{\f} _T^2 \hat{\f}_U^2
-\frac 1 4 \hat g _{MU}\hat{\f} _M^2 \hat{\f}_U^2
-\frac 1 4 \hat g _{MT}\hat{\f} _M^2 \hat{\f}_T^2\right], \nn
\eea
where we have included all possible quartic interactions invariant
under the
transformation $\hat{\f} _{U,T,M} \to -\hat{\f} _{U,T,M}$.
Here, $\hat{\f} _{U,T,M}$ denotes respectively
scalar fields with periodic (untwisted), antiperiodic
(twisted) and mixed boundary conditions.

Due to the above remark, the only constant classical background field
$\bac$ we may introduce is in the untwisted sector. Owing to this
structure, it is easily seen, that the analysis presented in the
previous sections is already enough to deduce in a very simple way the
effective potential for the theory (\ref{z2}).

The quantum potential consists now of three pieces, one coming from the
twisted, one from the untwisted and one from the mixed sector. The
interactions between the different types of scalar fields induce
different effective masses depending on the couplings involved. In
detail
\beq
V_4V_2 V(\Phi_U^2,\Phi_T^2,\Phi_M^2)
=V_4V_2[V_U(\Phi_U^2)+V_T(\Phi_T^2)+V_M(\Phi_M^2)],
                       \label{z3}
\eeq
with
\beq
V_4V_2V_i(\Phi_i^2)=\frac 1 2 \ln\left(-\frac{\Delta_i
+\Phi_i^2}{\kappa^2}\right),\quad i=U,T,M,
                              \label{z4}
\eeq
and with
\beq
\Phi_U^2=m_U^2+\frac 1 2 \lambda_U\back;\,\,\,
\Phi_T^2=m_T^2+\frac 1 2 g_{UT}\back;\,\,\,
\Phi_M^2=m_M^2+\frac 1 2 g_{UM}\back .           \label{z5}
\eeq
Similarly to Sect. 2, the four dimensional coupling constants
$\lambda_U$, $g_{UT}$ and $g_{UM}$, are related to the multidimensional
ones by $\lambda_U=\hat{\lambda}_U/volume(T^2)$, etc.

The contributions from the different sectors are calculated in the way
presented in Sect. 2, the relevant formula here is (\ref{a4}). In
order to express the result, we introduce
\beq
V_2V(\Phi^2;u_1,u_2) =
\frac{\pi^3L_1L_2}{12L^6}v^6\left[\frac{11} 6 -\ln\left(\frac{\Phi}
\kappa\right)^2\right] + H(\Phi^2;u_{1},u_{2}), \label{z6}
\eeq
where as before $v^2=(L\Phi/2\pi)^2$ and the second term is
defined by the formula (\ref{r4}) with $w_{1}=w_{2}=1$ (compare
with the analogous result (\ref{l7}) for the periodic case).
Then the regularized quantum potential reads
\beq
\vint =V(\Phi_U^2;0,0)+V(\Phi_T^2;1/2,1/2) +V(\Phi_M^2; 0,1/2).
                                    \label{z7}
\eeq
Imposing once more the renormalization conditions
(\ref{r21})-(\ref{r24}), the counterterms are easily
determined. From the previous discussion and the
comments in Sect. 3, it is clear that no additional calculations are
necessary. The only change being, that in eq.~(\ref{r5})-(\ref{r8}), the
terms $A(m_0^2)$, $B(m_0^2)$, $C^{(1)}(m_0^2,\la )$, $D^{(1)}(m_0^2,\la
)$, have to be replaced by a sum of three terms with the replacement
$m_0^2 \to m_U^2,m_T^2$ and $m_M^2$, respectively $\la \to
\lambda_U,g_{UT}$ and $g_{UM}$. Furthermore, for antiperiodic
respectively mixed boundary conditions, $H(\Phi^2;0,0)$ has to be
replaced by $H(\Phi^2;1/2,1/2)$, respectively by $H(\Phi^2;1/2,0)$. In
the same way, the effective potential is given by eq.~(\ref{r9}), once
$v^{(1)}(m_0^2,\la )$, given by eq.~(\ref{onep}), is replaced
by the sum of the
three contributions as described above. To have the $\lambda \bac ^{4}$-
model in the limit $L \rightarrow 0$ we take the renormalized constants
$\lambda_{U} = \lambda$, $m_{U}=m_{0}$. Furthermore $g_{UT}$,
$g_{UM}$ respectively $m_{T}$, $m_{M}$ are the renormalized
couplings of the interactions between the fields respectively masses
of the fields  with different boundary conditions, which we assume
to be all positive.

Representations similar to (\ref{vE-small})-(\ref{dv1-small})
in this case are (for $m_{0}^2 \neq 0$ and $\mu = 0$):
\[
v_{E}(x) = v_{M^{4}}(x) + \Delta v_{U}(x) + \Delta v_{T}(x)
   + \Delta v_{M}(x),
\]
where $\Delta v_{U}(x)$ is the same as in eq. (\ref{dv1-small}),
$\Delta v_{T}(x) = f(x,1/2,1/2,m_{T}^2/m_{0}^2,g_{UT}/
\lambda)$ and
$\Delta v_{M}(x) = f(x,1/2,0,m_{M}^2/m_{0}^2,g_{UM}/
\lambda)$, where the function $f(x,u_{1},u_{2},k,q)$ is given
by
\bea
\lefteqn{ f(x,u_{1},u_{2},k,q)  =  \frac{g}{\nu^2}
       \sum_{l_{1},l_{2} = -\infty}^{\infty}
       \left[  ((l_{1}-u_{1})^2  + (l_{2}-u_{2})^{2}
       + \nu (k+qx))^{2}    \right.} \nn  \\
        & & \times  \ln \frac{ (l_{1}-u_{1})^2  + (l_{2}-u_{2})^{2}
       + \nu (k+qx)}{(l_{1}-u_{1})^2  + (l_{2}-u_{2})^{2}
       + \nu k }
       - x \nu q ((l_{1}-u_{1})^2  + (l_{2}-u_{2})^{2}
       + \nu k)   \nn  \\
        & - & \left. \frac{3}{2} \nu^2 q^2 x^{2}
- \frac{\nu^3 q^3 x^{3} }
       {3((l_{1}-u_{1})^2  + (l_{2}-u_{2})^{2}
       + \nu k)} \right]  .
                                               \label{f-def}
\eea
Notice that $\Delta v_{U}(x)$ is equal to $f(x,0,0,1,1)$ but with the
term with $l_{1}=l_{2}=0$ being omitted from the sum. Scaling properties
of these functions, listed in Appendix C, show that the one-loop
contributions
have similar behaviours and varying values of $g_{UT}$, $g_{UM}$ and
the masses amounts to rescalings only. For this reason,
qualitatively
the effective potential $V_{E}(\bac)$ does not change and its properties
for small and large $\bac$ are the same as in the case of the
periodic field only.

However, the following important observation is in order. It is
seen, that in eq.~(\ref{z6}) the leading term for $\phi \to \infty$
does not depend on the boundary conditions imposed. Actually, it may be
shown that the leading behaviour of $v_E(x)$ for $\phi \to \infty$
is
\beq
v_E(x) \equiv -\frac 1 {6144\pi^3}\frac{\lambda_U}{m_U^4}\phi^6\left(
\lambda_U^3 +g_{TU}^3+g_{MU}^3\right).\label{asym}
\eeq
Thus one finds, that due to the interactions between the different
scalar fields the effective potential tends quicker to infinity than for
one scalar field only. This means, that the lifetime of the vacuum state
$\phi =0$, depending on the height and width of the potential barrier,
is reduced due to the presence of these interactions. In order to
reduce the lifetime
considerably one has to choose the coupling constants large enough.
However, at this point no further conclusions can be drawn, because for
these values one might enter into the strong coupling regime, where we
cannot
trust the one-loop result. We will come back to this point in the
next section, when dealing with stability in arbitrary dimensions.

\section{Stability of the theory in arbitrary dimensions}

Let us now consider the model of Sect. 2 in the $(4+n)$ dimensional
manifold $M^4\times T^n$ in order to analyze the dependence of the
stability on the number $n$ of compactified dimensions. Then,
instead of eq.~(\ref{zeta}) one is led to the zeta-function
\beq
\zeta(s;M^4 \times T^{n}) =
    \frac{\pi^2}{L^4}\left(\frac L {2\pi}\right)^{2s}
\frac{\Gamma (s-2)}{\Gamma (s)}Z_n^{v^2}(s-2;w_1,...,w_n),
                               \label{sta1}
\eeq
with $w_i=(L/L_i)^2$, $i=1,...,n$. In this general case it is still
possible to determine the analytical structure of the regularized
effective potential. The necessary analytical expressions are
summarized in Appendix B. One finds
\bea
V_n V(\Phi^2) &=&-\frac 1 4 \frac{\pi ^2}{L^4} \left\{{Z_n'}^{v^2}
(-2;w_1,...,w_n;u_1,...,u_n)\right.\label{sta2}\\
& & \left.\qquad +  Z_n^{v^2}(-2;w_1,...,w_n;u_1,...,u_n)
\left[\ln\left(\frac{L\kappa}{2\pi}\right)^2+\frac 3
2\right]\right\},\nn \eea
with
\beq
Z_n^{v^2}(-2;w_1,...,w_n;u_1,...,u_n) =\left\{
                   \begin{array}{cl}
             0, &  \mbox{for }\ n\  \mbox{odd},\\
             \frac{\dsp 2(-1)^{\frac n 2}\pi^{\frac n
2}v^{n+4}}{\dsp \sqrt{w_1...w_n}\left(\frac n 2 +2 \right)!},   &
 \mbox{for}\  n\     \mbox{even}.
        \end{array}\right.                  \label{sta3}
\eeq
Using (\ref{b3}) and (\ref{b4}), and the formula above, this
can be written more explicitly as follows:
for $n$ even
\bea
\lefteqn{   V_nV(\Phi^2) =\frac 1 2 (-1)^{\frac{n}{2}+1}\frac{\pi^{\frac
n 2 +2}}{\left(\frac n 2
+2\right)!}\frac{L_1...L_n}{L^{n+4}}v^{n+4}\times    }
                                          \label{sta4}\\
& &\qquad\qquad \left[\frac 3 2 +\sum_{k=3}^{\frac n 2 +2}\frac 1 k
-\ln\left(\frac{\Phi}{\kappa}\right)^2\right]\nn\\
& &-\frac{L_1...L_n}{L^{n+4}}v^{\frac n 2 +2}\sleln \exp\{2\pi
i[l_1u_1+...+l_nu_n]\}\times\nn\\
& &\qquad\wewn  ^{-\frac 1 2 \left(2+\frac n 2\right)}K_{\frac n 2
+2}\left(2\pi v \wewn^{\frac 1 2}\right)\nn
\eea
and for $n$ odd
\bea
\lefteqn{   V_nV(\Phi^2) =\frac 1 2 (-1)^{\frac{n-1} 2}
\frac{\pi^{\frac n 2 +3}}{\Gamma\left(\frac n 2
+3\right)}\frac{L_1...L_n}{L^{n+4}}v^{n+4}    }
                                      \label{sta5}\\
& &-\frac{L_1...L_n}{L^{n+4}}v^{\frac n 2 +2}\sleln \exp\{2\pi
i[l_1u_1+...+l_nu_n]\}\times\nn\\
& &\qquad\wewn  ^{-\frac 1 2\left(2+\frac n 2\right)}K_{\frac n 2
+2}\left(2\pi v \wewn^{\frac 1 2}\right).\nn
\eea
The leading asymptotic behaviour of the 1-loop
contribution to the effective potential before renormalization is:
for $n$ even
\beq
V_nV(\Phi^2) =(-1)^{\frac{n}{2}}\frac{\pi^{\frac
n 2 +2}}{2\left(\frac n 2
+2\right)!}\frac{L_1...L_n}{L^{n+4}}(v^2)^{n/2+2} \ln \frac{\lambda
   \back}{2 \kappa ^2}     \label{asymp1}
\eeq
and for $n$ odd
\beq
V_nV(\Phi^2) =(-1)^{\frac{n-1} 2}
\frac{\pi^{\frac n 2 +3}}{2\Gamma\left(\frac n 2
+3\right)}\frac{L_1...L_n}{L^{n+4}}(v^2)^{n/2+2},  \label{asymp2}
\eeq
where, as above, $v^{2} = (L\Phi)^2/(2\pi)^2$ and $\Phi^2 = m_{0}^2+
\lambda \back/2$. As we have seen in the previous sections,
in the case of the six-dimensional model the
renormalization does not change the leading behaviour of the effective
potential for $\phi\to \infty$. This is also true in the
$(4+n)$ dimensional space-time considered here. Indeed, for $n$ even,
in the renormalization conditions analogous to (\ref{r21})-(\ref{r24}),
the last one, which contains the highest derivative, would amount to
adding terms proportional to $\bac^{n+4}$. This does not change the
asymptotic behaviour (\ref{asymp1}). Now, let us turn to the case of
$n$ odd. As  is well known, for the $\lambda \phi^4$-theory in
$(4+n)$ dimensions, due to the $Z_{2}$-symmetry, the
only one-loop diagrams which are
ultraviolet divergent have a number of external legs
$\leq (n+3)$. Thus, for example, in five dimensions the one-loop
divergences are the same as in four dimensions. In this event the
renormalization conditions analogous to (\ref{r21})-(\ref{r24})
would amount to subtraction of terms up to $\bac^{n+3}$ and hence
would not affect the asymptotics (\ref{asymp2}).

Looking only
at the leading behaviour we find that stability or instability of
$V_{E}(\bac)$ for large
$\phi$ appears always pairwise.
For $k\in \nats$, one finds instability for the
dimensions $n=4k+2,4k+3$, and stability for $n=4k,4k+1$.

Let us now include once more interactions between the
scalar fields with different types of boundary conditions
existing on the spacetime $M^4\times T^n$. As is well
known, for different compactification lenghts $L_i$ there exist $2^n$
scalar fields on the spacetime $M^4\times T^n$. Assuming for simplicity,
that their couplings to the untwisted scalar field are approximately
equal, let us say $\lambda$, the leading behaviour, corresponding to
eqs.~(\ref{asymp1}) and (\ref{asymp2}), is multiplied by the factor
$2^n$. Thus even for small values of the coupling constant
$\lambda$, that is in the range
where our results are valid, we show that for
$n=4k+2,4k+3$, increasing dimensionality of the compact dimensions might
drastically reduce the lifetime of the vacuum state $\phi =0$. This
observation was our main reason for including interactions between
different types of scalar fields into our analysis.

\section{Conclusions}

We have calculated in this paper the effective potential for the
$\lambda
\phi^4$-theory on a $M^{4} \times T^{2}$ space-time and showed that it
is unbounded from below, namely that $V_{E}(\bac) \rightarrow -\infty$
when $\bac \rightarrow \infty$. The renormalization conditions
(\ref{r21})-(\ref{r24}) imposed guarantee that for small values of
$(\lambda \back L^{2})$, i.e. in the weak field limit or/and
for small size of the extra dimensions, the standard Coleman-Weinberg
potential is recovered. This is a manifestation of the
decoupling of heavy modes. The effect of decoupling in case of
the renormalized four-point Green function in the same theory
was studied in ref. \cite{decoupling}. We also have shown that the
asymptotics of the effective potential are not affected by the
renormalization procedure. We argue that it  will not change either
in any renormalization scheme based on adding counterterms
of minimal possible powers in $\bac$ to the classical
effective potential. In this sense the asymptotic behaviour of the
one-loop effective potential is a renormalization scheme
independent property of the theory.

The unboundness of $V_{E}(\bac)$ from below means that the
vacuum $\bac =0$ for $m_{0} \neq 0$ (Fig. 1) or the
non-trivial vacuum of the Coleman-Weinberg potential
for $m_{0}=0$ (Fig. 3) are metastable, if the
$\lambda \f ^4$-theory in four dimensions
is the low energy limit of the theory obtained by
dimensional reduction of the six-dimensional
one. Of course, it is true that if the size $L$ of the two
extra dimensions is small, the maximum of the potential
is high enough and the value of $\bac$ where $V_{E}=0$
lies far away, so it is natural to expect that
the lifetime of the vacuum is
very long. However, one can estimate that in the interval where $V_E\leq
0$ higher loop corrections are important and may change the result
drastically. Thus in the relevant range of values of $\phi$ one cannot
trust the one-loop result and further analysis will be necessary. A
possibility to tackle this problem would be to consider a
self-interacting $O(N)$ theory and to sum over all bubble graphs
dominant
for large $N$ in each order, thus obtaining reliable results at larger
values of $\phi$ \cite{onmodel}.

We saw that adding the sectors of fields with antiperiodic
and mixed boundary conditions on the torus $T^{2}$ and
interactions between them does not change the picture
qualitatively. However, we have seen that they probably reduce the
lifetime of the vacuum state.

We have also extended the analysis of the asymptotics of the
one-loop effective potential to the case of the spaces
$M^{4} \times T^{n}$ with any $n$. The result is that
in this approximation $V_{E}(\bac)$ is unstable for
$n=2,3 \pmod{4}$ and is stable for $n=0,1 \pmod{4}$.
When including interactions between scalar fields
with different boundary conditions,
our analysis suggests that for the cases $n=2,3$ (mod $4$)
contributions due to the compact space
$T^n$ may considerably reduce the lifetime of
the vacuum state $\phi =0$.

A few remarks are in order.
To complete
the analysis it would be important to see the effect of the
change of the subtraction point $\mu^2$ and to calculate the
$\beta$-function (similar to how it was done in ref.
\cite{coleman-weinberg})  and the running coupling constant
$\lambda (\mu)$. Then the position of the pole of the
latter defines the range of validity of $V_{E}(\bac)$.
In our case the running coupling constant due to contributions
of heavy modes has power-like terms in $\mu^2$ and the
position of the pole depends on $\nu$ (similarly to the
case studied in \cite{decoupling}). These issues
will be considered in a future paper.

The main messages of the result of the present analysis are
the following. First, we see that if a model appears as a
low-energy limit of some fundamental theory formulated on a
multidimensional space-time, like the superstring in ten
dimensions, then the issue of quantum stability of the vacua of this
model must be re-examined, taking into account contributions
of the tower of Kaluza-Klein modes. Second, if the vacuum
turns out to be metastable but $\nu = (m_{0}L)^2$
is small enough, then the lifetime of the vacuum is expected to be
rather long, in order to produce any physical effect.
This may be seen as a result of the very different magnitudes of the
scales involved, namely the inverse length $L^{-1}$ and the energy of
the particles. However, at the
early stage of the evolution of the Universe
this might change and $L$ could be comparable to the scale factor
of the 3-space of the Universe proportional to the inverse
temperature and the inverse energy scale of the particles
\cite{KK-cosmology}. Thus, decay of
the vacuum could be essential
and  give rise to new effects in Kaluza-Klein cosmology.
And the same could be also true for higher dimensions $n$, due to
possible interactions between different types of scalar fields.

\vspace{5mm}

\ni{\large \bf Acknowledgments}

It is a pleasure to thank G.~Kennedy, M.~Shaposhnikov, R.~Tarrach
and D.~Espriu for discussions and
remarks that turned into an improvement of the manuscript.
E.E. is gratefull to T. Muta and the whole Department of Physics,
Hiroshima University for interesting discussions and warm hospitality.
Yu.K.~and K.K. thank the Department ECM of Barcelona University for
warm hospitality.
This investigation has been supported by DGICYT (Spain), project No.
PB90-0022 and sabbatical grant
SAB 92 0267, and by CIRIT (Generalitat de Catalunya).
K.K. acknowledges financial support from the Alexander von Humboldt
Foundation (Germany).
\bs
\begin{appendix}
\renewcommand{\theequation}{{\mbox A}.\arabic{equation}}
\section{Appendix: Two dimensional Epstein-type zeta-functions}

\setcounter{equation}{0}

In this appendix we want to present some results on
Epstein-type zeta-functions necessary for the calculation
of the effective potential of the theory considered.

As we have seen, working with two compactified dimensions, one
is led to the functions
\beq
Z_2^{v^2}(\nu; w_1,w_2;u_1,u_2) =\suz [w_1(l_1-u_1)^2 + w_2(l_2-u_2)^2
+v^2]^{-\nu},    \label{a1}
\eeq
where different values of $u_i$ represent different boundary conditions
of the scalar field in the toroidal part $T^2$ of the manifold
$M^4\times T^2$.
In order to derive useful analytical representation of $Z_2^{v^2}$
we employ for $t\in \reals$, $z\in\komplex$ the re-summation formula
\cite{hille62}
\beq
\sum_{n=-\infty}^{\infty}\exp\{-tn^2+2\pi inz\} =\left(\frac{\pi}
t\right)\sum_{n=-\infty}^{\infty}\exp\left\{-\frac{\pi^2}t
(n-z)^2\right\},     \label{a2}
\eeq
which is nothing else but a rewriting of the famous Jacobi theta
function identity. Then one finds (for details see \cite{kir93})
\bea
\lefteqn{
Z_2^{v^2}(\nu;w_1,w_2;u_1,u_2)=\frac{\pi}{\sqrt{w_1w_2}} \frac{\Gamma
(\nu-1)}{\Gamma (\nu)} v^{2-2\nu}     }\label{a3}\\
& &+\frac{\pi^{\nu}}{\sqrt{w_1w_2}}\frac 2 {\Gamma (\nu)} \sun
\exp\left\{2\pi i [l_1u_1+l_2u_2]\right\}
  v^{1-\nu}\times\nn\\
& &\qquad\qquad\wew
^{\frac 1 2 (\nu -1)}K_{1-\nu} \left(2\pi v \wew ^{\frac 1 2}\right).\nn
\eea
This easily gives (with $w_i =(L/L_i)^2$)
\bea
\lefteqn{
Z_2^{v^2}(-2;w_1,w_2;u_1,u_2)=-\frac{\pi} 3 \frac{L_1L_2}{L^2}
v^6\left[\frac 1 3 -\ln v^2\right]     }\label{a4}\\
& &+\frac 4 {\pi^{2}}\frac{L_1L_2}{L^2}v^3\sun
\exp\left\{2\pi i [l_1u_1+l_2u_2]\right\}
 \wew ^{-\frac 3 2} K_{3} \left(2\pi v \wew ^{\frac
1 2}\right),\nn
\eea
which is the result needed for the calculation of the regularized
effective potential (\ref{l7}).

In order to see how the effective potential of Coleman and Weinberg
\cite{coleman-weinberg} is contained in our potential, we need the small
$v$ expansion of eq.~(\ref{a1}). It is given by \cite{kir93}
\bea
Z_2^{v^2}(\nu; w_1,w_2;u_1,u_2) &=& \left[
w_1u_1^2+w_2u_2^2+v^2\right]^{-\nu}\label{a5}\\
& &+\sum_{j=0}^{\infty}(-1)^j\frac{\Gamma(s+j)}{j!\Gamma (s)}Z_2
(s+j;w_1,w_2;u_1,u_2)v^{2j},\nn
\eea
with
\beq
Z_2(\nu; w_1,w_2;u_1,u_2) \sun
[w_1(l_1-u_1)^2+w_2(l_2-u_2)^2]^{-\nu}.\label{a6}
\eeq
This yields
\bea
\lefteqn{
{Z_2'}^{v^2}(-2;w_1,w_2;u_1,u_2) =-\frac 1 2
[w_1u_1^2+w_2u_2^2+v^2]^2\ln (w_1u_1^2+w_2u_2^2+v^2)  }\nn\\
& &+\frac 1 2 Z_2'(-2;w_1,w_2;u_1,u_2)+v^2\left[\frac 1 2
[w_1u_1^2+w_2u_2^2] +Z_2'(-1;w_1,w_2;u_1,u_2 )\right]\label{a7}\\
& &+\frac 1 2 v^4\left[\frac 3 2 +Z_2'(0;w_1,w_2;u_1,u_2)\right]-\frac 1
6 v^6 PP\,\,Z_2(1;w_1,w_2;u_1,u_2)\nn\\
& &+\sum_{j=2}^{\infty}(-1)^j\frac{(j-1)!}{(j+2)!} v^{2j+4}
Z_2(j;w_1,w_2;u_1,u_2),\nn
\eea
with $PP\,\,Z_2( 1;w_1,w_2;u_1,u_2)$ denoting the finite part of
$Z_2(1;w_1,w_2;u_1,u_2)$.
Eq. (\ref{a7}) is the result needed for the derivation of (\ref{vier1}).

\renewcommand{\theequation}{{\mbox B}.\arabic{equation}}

\section{Appendix: Epstein-type zeta-functions in arbitrary dimensions}

\setcounter{equation}{0}
Compactifying $n$ dimensions, one is naturally led to the functions
\beq
Z_n^{v^2}(\nu ;w_1,...,w_n;u_1,...,u_n)=\slel [w_1(l_1-u_1)^2+...+
w_n(l_n-u_n)^2+v^2]^{-\nu}.\label{b1}
\eeq
Here the analytical continuation reads
\bea
\lefteqn{
Z_n^{v^2}(\nu;w_1,...,w_n;u_1,...,u_n)=\frac{\pi^{\frac
n 2}}{\sqrt{w_1...w_n}}
\frac{\Gamma (\nu-\frac n 2)}{\Gamma (\nu)} v^{n-2\nu}     }\label{b2}\\
& &+\frac{\pi^{\nu}}{\sqrt{w_1...w_n}}\frac 2 {\Gamma (\nu)} \sleln
\exp\left\{2\pi i [l_1u_1+...+l_nu_n]\right\}
 v^{\frac n 2-\nu}\times\nn\\
& &\qquad\qquad\wewn
^{\frac 1 2 (\nu -\frac n 2)}K_{\frac n 2-\nu} \left(2\pi v \wewn
^{\frac 1 2}\right). \nn
\eea
Thus we find the quantities relevant for the effective potential. For
$n$ even, we have ($w=(L/L_i)^2$)
\bea
\lefteqn{  \hspace{-2.5cm}
{Z_n'}^{v^2}(-2;w_1,...,w_n;u_1,...,u_n) =\frac{2(-1)^{\frac n
2}}{\left(\frac n 2 +2\right)!}\frac{\pi^{\frac n
2}L_1...L_n}{L^n}v^{n+4}\left[\psi\left(\frac n 2
+3\right)-\psi(3) -\ln v^2\right]     }\nn\\
& &\hspace{-2.5cm}+\frac 4 {\pi^2}\frac{L_1...L_n}{L^n}v^{\frac n 2
+2}\sleln \exp\left\{2\pi i [l_1u_1+...+l_nu_n]\right\}
\times\label{b3}\\
& &\hspace{-2.5cm}\qquad\qquad\wewn
^{-\frac 1 2 (2+\frac n 2)}K_{\frac n 2 +2} \left(2\pi v \wewn
^{\frac 1 2}\right), \nn
\eea
whereas for $n$ odd we find
\bea
\lefteqn{
{Z_n'}^{v^2}(-2;w_1,...,w_n;u_1,...,u_n)=(-1)^{\frac{n+1} 2}
\frac{2\pi^{\frac n 2 +1}L_1...L_n}{\Gamma\left(\frac n 2 +3\right)
L^n}v^{n+4}
    }\nn\\
& &+\frac 4 {\pi^2}\frac{L_1...L_n}{L^n}v^{\frac n 2 +2}\sleln
\exp\left\{2\pi i [l_1u_1+...+l_nu_n]\right\}
\times\label{b4}\\
& &\qquad\qquad\wewn
^{-\frac 1 2 (2+\frac n 2)}K_{\frac n 2 +2} \left(2\pi v \wewn
^{\frac 1 2}\right). \nn
\eea
Eq.~(\ref{b3}) and (\ref{b4}) are the needed results to analyze the
large $\phi$ behaviour in the self-interacting theory in a
spacetime of the form $M^4\times T^n$.

\renewcommand{\theequation}{{\mbox C}.\arabic{equation}}

\section{Appendix: One-loop contributions to the effective potential}

\setcounter{equation}{0}

For $m_{0} \neq 0$ the terms $\Delta v_{2}(x)$ and
$\Delta v_{3}(x)$ in eq. (\ref{vE-small}) are
\bea
\Delta v_{2}(x) & = & \frac{g x^2}{2} \sun \hspace{3mm}\left\{ -2 \ln
\frac{
         N^{2}(l_{i})+\nu(1+y)}{N^{2}(l_{i})+\nu} + \frac{2y\nu}
         {N^{2}(l_{i})+\nu(1+y)}    \right.         \label{c1} \\
         & - & \left. \frac{172}{3} \frac{y^2 \nu^2}{[N^{2}(l_{i})
 +\nu(1+y)]^2}
 + \frac{80y^3 \nu^3}{[N^{2}(l_{i})+\nu(1+y)]^3} -
 \frac{32 y^4 \nu^4}{[N^{2}(l_{i})+\nu(1+y)]^4}  \right\}
                                         \nn
\eea
and
\bea
\Delta v_{3}(x) & = & \frac{2 g \nu x^3}{3} \sun\hspace{3mm} \left\{
           \frac{1}{N^{2}(l_{i})+\nu} - \frac{1}{N^{2}(l_{i})+\nu(1+y)}
                          	    \right. \label{c2} \\
  & + & \left. \frac{6 y \nu}{[N^{2}(l_{i})+\nu(1+y)]^2}
            - \frac{8 y^2 \nu^2}{[N^{2}(l_{i})+\nu(1+y)]^2}+
           \frac{16}{5} \frac{y^3 \nu^3}{[N^{2}(l_{i})+\nu(1+y)]^4}
  \right\},       \nn
\eea
where
\beq
 N^{2}(l_{i}) = l_{1}^2 w_{1} + l_{2}^2 w_{2}.   \label{c3}
\eeq

For $m_{0} = 0$ the quantum corrections in eq. (\ref{uE-small})
are equal to:
\bea
\Delta u_{1}(z) & = & 6g \sun\hspace{3mm} \left\{ \frac{[N^{2}(l_{i}) +
zt]^2}
          {t^2} \ln \frac{N^{2}(l_{i}) + zt}{N^{2}(l_{i})} - \frac{z}{t}
  N^{2}(l_{i})        \right.        \nn  \\
         & - & \left. \frac{3}{2} z^2 - \frac{1}{3}
 \frac{z^3 t}{N^{2}(l_{i})}
          \right\},         \label{c4}   \\
\Delta u_{2}(z) & = & 3gz^2 \sun \hspace{3mm} \left\{ -2
          \ln \frac{N^{2}(l_{i}) + t}{N^{2}(l_{i})} + \frac{2t}
  {N^{2}(l_{i}) + t} - \frac{172}{3} \frac{t^2}
  {[N^{2}(l_{i}) + t]^2}
               \right.        \nn  \\
    & + & \left. \frac{80 t^3}{[N^{2}(l_{i}) + t]^3}  -
          \frac{32 t^4 }{[N^{2}(l_{i}) + t]^4}
          \right\},         \label{c5}   \\
\Delta u_{3}(z) & = & 12gt^2 z^3 \sun \hspace{3mm}\left\{
\frac{2}{N^{2}(l_{i})
          (N^{2}(l_{i}) + t)} + \frac{6}{[N^{2}(l_{i}) + t]^2}
          \right.        \nn  \\
    & - & \left. \frac{8t}{[N^{2}(l_{i}) + t]^3} +
          \frac{16}{5} \frac{t^2}{[N^{2}(l_{i}) + t]^4}
          \right\},         \label{c6}
\eea
where $N^{2}(l_{i})$ is given by eq. (\ref{c3}).

The function $f(x,\nu,u_{1},u_{2},k,q)$ defined in eq. (\ref{f-def})
(here we indicate its dependence on $\nu$ explicitly),
which characterizes one-loop contribution of the heavy
modes with arbitrary boundary conditions, posesses the
following scaling properties:
\[  f(x,\nu,u_{1},u_{2},k,\lambda q) = f(\lambda x,\nu,u_{1},u_{2},k,q), \]
\[ f(x,\nu,u_{1},u_{2},\lambda k,q) = \lambda^2
            f(x/\lambda,\lambda \nu,u_{1},u_{2},k,q).   \]
These scaling properties have been used to show explicitly that
interactions between the
different types of scalar fields (see section 4) do not qualitatively
change the effective potential.

\end{appendix}

\newpage

\section*{Figure captions}

\begin{description}
  \item[Fig. 1] Plots of the functions $v_{M^{4}}(x)$ (labelled
                by CW) and
                $v_{E}(x)$ for the case of a massive field
$\bac$ with periodic boundary
conditions; $y=0$, $\xi=0$ and $w_1=w_2=1$ for
various $\nu$. The number near the curve
corresponds to the value of $\nu$.
  \item[Fig. 2] Plots of the functions $u_{M^{4}}(z)$ (labelled
                by CW) and $u_{E}(z)$
                for the case of a massless field $\bac$ with
periodic boundary conditions; $\eta = 0$ and
$w_1=w_2=1$. The number near the curve
corresponds to the value of $t$.
  \item[Fig. 3] The same plots as in Fig. 2 but for the interval
                $0 < z < 50$.
\end{description}

\end{document}